\magnification=\magstep1
\settabs 18 \columns

\input epsf
 
\hsize=16truecm

\def\b{\bigskip}
\def\bb{\bigskip\bigskip}

\def\no{\noindent}
\def\r{\rightline}
\def\ce{\centerline}
\def\ve{\vfill\eject}

\def\r{\rightline}

\def\harr#1#2{\smash{\mathop{\hbox to .25 in{\rightarrowfill}}
 \limits^{\scriptstyle#1}_{\scriptstyle#2}}}

\def\R{{\cal R}}

\def\today{\ifcase\month\or January\or February\or March\or April\or
May\or June\or July\or
August\or September\or October\or November\or  December\fi
\space\number\day, \number\year }

\r {Entropy}

\r \today
\bb\bb\bb

\def\Rrm{\hbox{\rm I\hskip -2pt R}}

\def\e{\rm e}

\def\p{\partial}
 
\def\sqr#1#2{{\vcenter{\vbox{\hrule height.#2pt
\hbox{\vrule width.#2pt height#2pt \kern#2pt
\vrule width.#2pt}
\hrule height.#2pt}}}}

 \def\1/2{{\scriptstyle{1\over 2}}}
 \def\a/2{{\scriptstyle{3\over 2}}}
 \def\5/2{{\scriptstyle{5\over 2}}}
 \def\7/2{{\scriptstyle{7\over 2}}}
 \def\3/4{{\scriptstyle{3\over 4}}}

\font\steptwo=cmb10 scaled\magstep2

\magnification=\magstep1

\def\sqr#1#2{{\vcenter{\vbox{\hrule height.#2pt
\hbox{\vrule width.#2pt height#2pt \kern#2pt
\vrule width.#2pt}
\hrule height.#2pt}}}}

\def \r{\rightarrow}

\b

\def\picture #1 by #2 (#3){
  \vbox to #2{
    \hrule width #1 height 0pt depth 0pt
    \vfill
    \special{picture #3}  
    }
  }

\def\scaledpicture #1 by #2 (#3 scaled #4){{
  \dimen0=#1 \dimen1=#2
  \divide\dimen0 by 1000 \multiply\dimen0 by #4
  \divide\dimen1 by 1000 \multiply\dimen1 by #4
  \picture \dimen0 by \dimen1 (#3 scaled #4)}
  }

\parindent=1pc

  \magnification=\magstep1
\settabs 18 \columns
 
\input epsf

\hsize=16truecm

\def\b{\bigskip}
\def\bb{\bigskip\bigskip}

\def\no{\noindent}
\def\r{\rightline}
\def\ce{\centerline}
\def\ve{\vfill\eject}

\def\r{\rightline}

\def\harr#1#2{\smash{\mathop{\hbox to .25 in{\rightarrowfill}}
 \limits^{\scriptstyle#1}_{\scriptstyle#2}}}

\def\R{{\cal R}}

\def\today{\ifcase\month\or January\or February\or March\or April\or
May\or June\or July\or
August\or September\or October\or November\or  December\fi
\space\number\day, \number\year }

\def\Rrm{\hbox{\rm I\hskip -2pt R}}

\def\e{\rm e}

\def\p{\partial}
 
\def\sqr#1#2{{\vcenter{\vbox{\hrule height.#2pt
\hbox{\vrule width.#2pt height#2pt \kern#2pt
\vrule width.#2pt}
\hrule height.#2pt}}}}

 \def\1/2{{\scriptstyle{1\over 2}}}
 \def\a/2{{\scriptstyle{3\over 2}}}
 \def\5/2{{\scriptstyle{5\over 2}}}
 \def\7/2{{\scriptstyle{7\over 2}}}
 \def\3/4{{\scriptstyle{3\over 4}}}

\font\steptwo=cmb10 scaled\magstep2

\magnification=\magstep1

\def\sqr#1#2{{\vcenter{\vbox{\hrule height.#2pt
\hbox{\vrule width.#2pt height#2pt \kern#2pt
\vrule width.#2pt}
\hrule height.#2pt}}}}

\def \r{\rightarrow}

\b

\def\picture #1 by #2 (#3){
  \vbox to #2{
    \hrule width #1 height 0pt depth 0pt
    \vfill
    \special{picture #3}  
    }
  }

\def\scaledpicture #1 by #2 (#3 scaled #4){{
  \dimen0=#1 \dimen1=#2
  \divide\dimen0 by 1000 \multiply\dimen0 by #4
  \divide\dimen1 by 1000 \multiply\dimen1 by #4
  \picture \dimen0 by \dimen1 (#3 scaled #4)}
  }

\parindent=1pc

\smallskip
  \pageno=1

  {\ce {\steptwo  On entropy in eulerian thermodynamics} }
\b

\ce{ C. Fronsdal and A. Pathak}

\ce{\it Department of Physics and Astronomy, University of Californa Los Angeles}
\b
\no{\it ABSTRACT} To the student of thermodynamics the most difficult subject is entropy. In this paper we examine the actual, practical application of  entropy to
two simple systems, the homogeneous slab with fixed boundary values of the temperature, and an isolated atmosphere in the presence of the static gravitational field.
The first gives valuable insight into the nature of entropy that is subsequently applied to 
the second system.

  It is  a basic tenet of thermodynamics  that the equilibrium of an extended, homogeneous and isolated system is characterized by a uniform temperature distribution and it is a  strongly held belief that this
remains true in the presence of gravity. We find that this  is consistent with the equations of extended thermodynamics but that entropy enters in an essential way.
The principle of equivalence takes on a new aspect.
 \b
\no{\bf I. Introduction}

To avoid misunderstanding we must  take care to define our context, as we have tried
to do in the title of this paper; namely, thermodynamics as it is developed in textbooks,
including the local extension  - including hydrodynamics - that  is sometimes referred to as extended thermodynamics,
a eulerian field theory in which the dynamical variables are temperature, density and pressure (densities and pressures), that is responsible for the description of equilibrium
configurations, as well as dynamical phenomena such as the propagation of pressure waves. Insight  derived from  kinetic and atomic theory is essential, but it will be used only after
translation to thermodynamic field theory.

Because entropy is a difficult concept,  we shall narrow the context to the comparatively simple special case of ideal gases. Experiments carried out in the beginning of the 19'th century    led to the discovery  of power laws   (Poisson 1835) that relate density, temperature and pressure, 
$$
p\propto \rho^\gamma,~~ \rho \propto T^n,~~ \gamma = 1 + {1\over n},\eqno(1.1)
$$
valid under conditions of thermal isolation and equilibrium, in the absence of external forces.  This says that $\rho/T^n$ remains unchanged for an adiabatic process. \footnote*{For example, it remains uniform and constant during sound propagation  (Laplace...).} A perfect gas is one for which these laws are exact; for such gases it has been shown that the specific entropy can be expressed, up to an inessential additive constant, as
$$
S =- \R \ln{\rho\over T^n},\eqno(1.2)
$$
where $n$, as in Eq.(1), is the adiabatic index of the gas. The value of $n$ for a specific gas can be derived from kinetic theory, but in our context it is a given datum. 

 In the case of
an equilibrium configuration of a thermally isolated ideal gas, in the absence of external forces,  all the fields are  uniform, and the entropy is reduced to a real number
that effectively parameterizes the adiabats of the system.  

Questions that relate to changes in the entropy have to be phrased with care. In this paper we shall discuss entropy within the context of the equations of motion, for
stationary configurations.

\b
The equations that are used to describe an  isolated, one component  thermodynamic system are the equation of continuity, the Navier-Stokes 
equation and energy conservation. In the simpler case
when dissipation and shear forces are neglected they reduce to conservation laws:
the equation of continuity (conservation of mass), the Bernoulli equation (conservation of momentum) and energy conservation. To this must be added an equation of state.
Here we are going to restrict the attention to equilibrium configurations, for which all the fields are time independent and the velocity field is zero. The full list of equations is then the hydrostatic equation
$$
{\rm grad} ~p  = 0,\eqno(1.3)
$$
and the equations of state that in the case of an ideal gas is  the gas law
$$
p = \R \rho T\eqno(1.4)
$$
and the expression for the internal energy density
$$
u = n\R  \rho T. \eqno(1.5)
$$
The polytropic relations can be deduced from the last 2 equations if, in addition, one
assumes that the specific entropy is uniform. \footnote*{ Here we assume that the entropy density $s = \rho S$, hence attributable to the gas. } In general the set (1.3-5) must be supplemented with some statement about the entropy.
To understand the role of entropy it is necessary to go to the foundations of thermodynamics.

\ve 
 The laws of equilibrium thermodynamics of a homogeneous substance, in the absence of external forces, can be summarized as a Gibbsean variational principle,
$$
\delta H =0,~~H = F(T,V) + ST + PV,
$$
where $F$ is the free energy and $V$ is the volume. In the case of an ideal gas
$$
F(T,V) = -\R T\ln(V T^n).\eqno(1.6)
$$
Here  $T,V,S$ and $P$ are independent dynamical variables, but only $T$ and $V$
are varied, to give the two principal relations
$$
{\delta  H\over \delta T}\big|_{V,S,P} = 0,~~{\rm that~ is,~} ~{\p F\over \p T} \big|_V+ S = 0,\eqno(1.7)
$$
and
$$
 {\delta H\over \delta V}\big|_{T,S,P} =0;~~{\rm that~ is,~}~ {\p F\over \p V}|_T+ P = 0.\eqno(1.8)
$$
Given the structure of the action, there can be no question of variation with respect to $S$ or $P$; these are parameters that distinguish one equilibrium  configuration from another, held fixed while $T$ and $V$ are varied.

 The equations obtained by minimizing the action are equivalent to those in the standard formulation of thermodynamics in the Helmholtz representation (Legendre transform of energy representation) with the entropy always held fixed. \footnote*{
``It is perhaps superfluous at this point to stress again that thermodynamics is logically complete within either the entropy or the energy representations and self-contained within either the entropy or the energy representations and that the introduction of the transformed representations is purely a matter of convenience."
 (Callen , 1985, page   )}

In the case of an ideal gas the first of the two equations, Eq. (1.7),  gives the formula 
$$
S = \R \ln(V T^n) +n\R
$$
 for the entropy. 
Contrary to the impression that may be obtained from some texts, this formula is not 
primarily to be understood as
a means of evaluating or defining the entropy for a given thermodynamical configuration,  but as an equation of motion or  
constraint that is to be satisfied by the dynamical variables $T, \rho$, depending on 
an {\it a priori} choice of $S$. The dynamics that results from a particular choice of entropy is referred to as ``adiabatic".

The local extension of the theory is far from being a trivial matter.  The dynamical variables are fields over $\Rrm^3$, $F$ and $S$ are interpreted as specific densities
(per gram or per particle) and $V = 1/\rho$ is now the specific volume.  The total energy  is
$$
H = \int  d^3 x \Big(\rho\,\vec v\,^2/2 +f(T,\rho) + sT \Big)+ PV,\eqno(1.9)
$$
where
$$
f(T,\rho) = \rho F(T,V),~~ \rho = 1/V.
$$
and $s$ is an entropy density. It is this last variable that is the center of interest of this paper. \underbar{Provisionally} we suppose that
 $$
 s = \rho S,\eqno(1.10)
 $$
 where the specific entropy $S$ is an independent  variable.  Both $P$ and $S$ are parameters  together their values  characterize the adiabatic families of configurations. That is not to say that these parameters can be chosen at random, as we shall see. The dynamical variables are now the fields $T, \rho$ and $S$, 
 while $P$ enters only as the pressure at the boundary.  \footnote *{ When the focus is exclusively on equilibrium configurations it may be said that independent variation of $s$ (with only the total entropy $\int d^3 x \,s\,$ held fixed)  leads to the desired result that $T$ is uniform, but in the general dynamical context this cannot be true; hence variation of the entropy is not to be considered in the present context.} \break
 
 Variation of  $H$ with respect to $T$ and $\rho$, with $S$ and $P$ held fixed,  gives  local versions of Eq.s (1.7-8). Variation of $T$:
 $$
 {\p f\over \p T} + s = 0.\eqno(1.11)
 $$
In the case of an ideal gas 
$$
f(T,\rho) = \R \rho T \ln{\rho\over T^n} \eqno(1.12)
$$
and (1.11) becomes, under the assumption (1.10),
$$
\R  (\ln{\rho\over T^n}-n)  +  S = 0;\eqno(1.13)
$$
The specific entropy $S$ is a field, in principle arbitrary, that is held fixed under the variation of $T$ and $\rho$. An adiabatic motion is one that takes place with no change in the value of this field; that is, a motion for which $S$ is independent of the time. In the case that $S$ is uniform, the last equation is the  polytropic condition $\rho \propto T^n$ and the numerical value of $S$ then parameterizes the family of adiabats.
\footnote {**}{The expression for the free energy density can be derived from the two equations of state that characterize an ideal gas, the gas law and the expression for the $u = n\R T$ for the internal energy. The expression for the specific entropy density
then follows from this condition, that the hamiltonian (1.9) be extremal with respect to variations of the temperature. The fact that the same was derived by kinetic theory is evidence of the 
compatibility between kinetic theory with thermodynamics, if such were needed.}

  In the general case, local variation of the density, vanishing at the boundary, with the total mass $\int d^3x\, \rho$ held fixed gives, for equilibrium configurations ($\vec v = 0$ and all fields time independent),
$$
{\rm grad}\big({\p f\over \p \rho} + T{\p s\over \p\rho}\big) = 0,~~ {\rm or}~~  {\p f\over \p \rho} + T{\p s\over \p \rho} 
={\rm constant}.\eqno(1.14)
$$
It reduces to the hydrostatic condition ${\rm grad} \,p= 0$  under a certain proviso.
The definition of the thermodynamic pressure is 
$$
p=-{\p F\over \p V} = \rho^2{\p\over\p \rho}{ f\over\rho} = (\rho{\p\over \p \rho}-1)f,\eqno(1.15)
$$
Using the last 3 equations  and (1.11) we find \footnote {***}{ The calculation: We have
$
{\rm grad }\, p= \rho\,{\rm grad}\,  {\p f \over\p\rho} + {\p f \over \p\rho}{\rm grad}\, \rho - {\rm grad }\, f,
$
and
$
{\rm grad}\, f = {\p f\over \p \rho}{\rm grad}\,\rho + {\p f\over \p T}{\rm grad T}.
$
 Add these two and use (1.11) and (1.14 to get (1.16).}
$$
{\rm grad }\,p= [(1-\rho{\p\over \p \rho})s]\,{\rm grad}\, T-\rho T\,{\rm grad} {
\p s\over \p\rho}.\eqno(1.16)
$$
\b
In the case that $s = \rho S$, with $S$ uniform (and  independent of $\rho$), this leads to the simple statement (1.3), 
$$
{\rm grad} \, p = 0.
$$
This is the hydrostatic equation,  the Bernoulli equation specialized to equilibrium configurations, in the absence of external forces.
 
In the case of an ideal gas, the pressure defined by (1.15) is  $p=\R\rho T$ and the internal energy
density $f+ sT$ reduces, in view of (1.13), to $n\R \rho T$. 

The set of variations that vanish at the boundary is supplemented by variations  of density and volume,   the total mass  held fixed. These variations yield the additional result
that the thermodynamical pressure agrees with the external pressure $P$;
this is what justifies the interpretation of the quantity (1.15) as pressure. 
\b
The set of equations 
 (1.3-5) is thus in full accord with thermodynamics, in the case that $s = \rho S$, with $S$ uniform and  independent of $\rho$. Going beyond the special case of an ideal gas we can
always replace the equations of state by an expression for the free energy density.
But to justify the hydrostatic condition it is necessary to place the following condition on the entropy:
$$
[(1-\rho{\p\over \p \rho})s]\,{\rm grad}\, T-\rho T\,{\rm grad} {
\p s\over \p\rho}=0.\eqno(1.17)
$$
  The traditional approach respects the structure of (extended) thermodynamics only to the extent that this is true.  In the polytropic case all the fields are necessarily uniform in  stationary configurations. The applications that we shall study in this paper  are not of that kind; the entropy plays a more interesting role.
 \b
We shall try to advance our understanding of entropy by a careful examination of some simple systems for which   the homogeneity of the equilibrium
configurations, and the general dynamics, is disturbed by external circumstances,
expressed as boundary conditions or as external forces, which shall compel
us to allow  for a non uniform entropy and thus to learn how the hydrostatic equation can be justified.

Our first example, in Sections II and III,  deal with the stationary conduction of heat through a slab, with the two plates maintained at constant, unequal temperature. The other example, in Section IV,  examines the effect of gravity on an isolated column in equilibrium under the assumption that the temperature, at equilibrium, is uniform.
  \b
  
  {\bf Remark.} That the functional (1.9) is a fully fledged hamiltonian is seen as follows.
Introduce the Poisson bracket
$$
\{A,B\} = \int d^3x\big({\p A\over \p v_i}\p_i{\p B\over \p \rho}- {\p B\over \p v_i}\p_i{\p  A\over \p \rho}   \big) ;
$$
then 
$$
\dot \rho = \{H, \rho\} = -\p_i(\rho v_i),
$$
$$
\dot v_i = \{H,v_i\}=  -\p_i{\p h\over \p \rho}
$$
and, of course, $\dot H = \{H,H\} = 0$.   

\b\b  Ê	
\no{\bf  II. First example.  The slab}

We consider  the case of an ideal gas that is confined between two parallell (horizontal) plates that are maintained at constant  and uniform temperatures $T_1 < T_2$,
respectively. It is required that the temperature field satisfy the boundary conditions that make it continuous across the boundary; in particular, it can not be uniform.  One can envisage a  stationary configuration, not in thermal equilibrium, in which
the velocity field is zero and the dynamical variables are time independent. 

The salient point that attracts attention to this simple example is the fact that the boundary conditions are incompatible with a uniform temperature. Since we are 
definitively dealing with an ideal gas we must use the   expression (1.6) for the free energy density 
$$
f(T,\rho) = \R \rho T\ln{\rho\over T^n}; \eqno(2.1) 
$$
the thermodynamical pressure is thus $p = \R\rho T$ and the formula (1.2) for the specific entropy of the gas applies. In the absence of flow the  Bernoulli equation implies that $p$ must be uniform. Since the temperature cannot be uniform because of the boundary conditions, neither is the density. Then it follows from formula (1.2)  that  the specific entropy  density is not uniform. \footnote*{Here we are still assuming that Eq.(1.10) holds  , $s = \rho S$, but see below.}

The two fundamental relations of thermodynamics are Eq.s (1.11) and (1.14). For an ideal gas they are
$$
\R\rho(\ln k - n) + s = 0,~~{\rm and}~~ \R T(\ln k + 1) + {\p s\over \p \rho}T = c = {\rm constant},\eqno(2.2)
$$
where $k := \rho/T^n$.
Multiplying the first by $T$ and the other by $\rho$ and subtracting the first from the second we obtain
$$
[(n+1)\R\rho  + (\rho{\p s\over \p \rho}-s) ] T= c \,\rho.
$$
This shows that, if $s$ is linear in $\rho$ -  Eq.(1.10) with $S$ independent of $\rho$ -   then the temperature is necessarily uniform,
contradicting the boundary conditions. This conclusion is reached irrespective of 
any assumptions about the specific entropy density; letting it vary from point to point does not help.

This calls for a generalization of the assumption (1.10) and the simplest possibility is to add an extra term to the density
$$
s = \rho S + S_{\rm ex},\eqno(2.3)
$$
with both $S$ and $S_{\rm ex}$ independent of the dynamical variables. Then Eq.s (2.2) become
$$
\big( \R (\ln k + 1) + S)T = c\R,~~{\rm and}~~[(n+1)\R  -S_{\rm ex} /\rho] T= c\R,\eqno(2.4) 
$$
 These two equations give the density and the temperature for any choice of the distributions $S$ and $S_{\rm ex}$.   The boundary conditions restrict the choice;
 since $T$ is not uniform, $S_{\rm ex} \neq 0$.
 \b
Equation (1.3), ${\rm grad}\, p = 0$,  demands that  $p$   be  uniform.   We can reconcile this with thermodynamics by demanding cancelation between the two terms on the right hand side of Eq.  (1.16),
$$
S_{\rm ex}\,{\rm grad} \,T =\rho T\,{\rm grad} \,S.\eqno(2.5)
$$
This makes the pressure uniform. Conversely, if $p = \R \rho T$ is uniform, then
this equation is a consequence of (2.4). In this case, since $T$ is not uniform neither is $S$.\b
{\it Summary.} Every distribution of density and temperature is possible, but  when the present boundary conditions apply,  then  $S_{\rm ex} \neq 0$ and $S$ is not uniform.  There are two interesting special cases:
\b1.  Maintain ${\rm grad}~ p = 0$,  accept that the specific entropy is not uniform.

2. Insist that the specific entropy be uniform, then  ${\rm grad}\, p  = S_{\rm ex}{\rm grad}\,T \neq 0 $. 

\vskip-1cm 
$$ \eqno(2.6) $$
\b 
The main conclusion is  that  the traditional treatment of the slab is consistent with thermodynamics, but that  it requires  a contribution to the entropy that is not linear in the density. The  question  remains  whether this is allowed by the 
interpretation of entropy, as understood in statistical mechanics.
We think that the answer is affirmative and attempt to justify this as follows.
 
The total entropy  density is $s = \rho S + S_{\rm ex}$. The term $S_{\rm ex}$ is the entropy of the ``background", that enters the 
calculations only through the boundary conditions. The implication of this is that
this part of the entropy stands for an interaction of the system under investigation with the backgound.
 The dynamics of the background is trivialized in this case by
ignoring any effect on it that is due to the contact with our system.  

The system (1.3-1.5) is incomplete. In the simplest case it is supplemented by the
statement that the specific entropy is uniform and  $S_{\rm ex} = 0$. In general it  must be supplemented by some information about the entropy. Once the entropy is specified   Eq.s (2.4) are the equations of motion.  These 2 equations come from (2.2), where the first is a modified polytropic relation and the second is an integrated form of the hydrostatic condition  (1.16).  Both are fundamental equations of thermodynamics, the local extensions of (1.7-8).   If the hydrostatic equation is maintained  in the original form (${\rm grad} \, p = 0$)  then, for any choice of the entropy,  Eq.(2.5) must be respected.
\b\b

\no{\bf III. Mechanical model of the slab}

\ce{\it  Interpretation and clarification of the $\rho S _{\rm ex} $ form for entropy density}  

Let us summarize the notation:  $S$ is the  specific entropy of the gas, $s$ the total entropy density, $V$ is the total volume  and $N$ is the number of particles. Let $S_T$ be the total measured entropy. $s = S_T/ V$. In our considerations of the slab we had to introduce the extra variable $S_{\rm ex} $ . This implies  that we are postulating an extra degree of freedom (call it $\eta$). Now $s$ will scale correctly only if $\eta$ is also scaled with $U, V$ and $N$.\footnote* {
Next sentence is a repetition. Suggest to omit it .}  In the simplest assumption, which is made in this paper, we write the measured entropy as $(S_T - S_{\rm ex}) + S_{\rm ex}$, where $(S_T - S_{\rm ex})$ scales with ($U,V,N$) while $S_{\rm ex}$ in general scales with ($U, V, N, \eta$).

In order to clarify the interpretation of the dynamical equations, \footnote{** }{ Omit: that were obtained above upon demanding that the total energy is stationary under variation of $\rho$ and $T$ with both  $S$ and $S_{\rm ex}$ held constant}  we briefly take a closer look at the form of the energy $H$ that appears in the action and compare it with the expression for $U$ (the internal energy) in the standard Helmholtz formulation. The free energy of the gas is defined as the Legendre transform of its energy, $U_{\rm gas}$, w.r.t. its entropy $S_{\rm gas}$, $S_{\rm gas} = V\rho S$.  We have: $F_{\rm gas}(T,\rho) = U_{\rm gas}-TS_{\rm gas}$, so that $U_{\rm gas} = F_{\rm gas}+TS_{gas}$ is the entire contribution of  the ideal gas to the total energy {H}. The other term in {H} is then the contribution of an extra system, always at the same temperature as the gas. $U_{\rm ex} =F_{\rm gas}+ TS_{\rm ex}$ is the only part of its energy that is varied during any thermodynamic process. Because the only work delivered by the extradegree of freedom $\eta$  to a reversible work source is $-TdS_{\rm ex}$ and because its temperature is the same as the gas throughout, it may be interpreted as a thermal reservoir as defined in the familiar standard formulations: The 2nd order fluctuations of the extensive variables are zero.  \footnote{***}{Omit the next sentence.}Alternately, the gas may be a reservoir for this extra system. The bottom line is that the work ($W$) delivered to a reversible work source is given as:
$dW = -dU_{\rm gas} -dU_{\rm ex}=   -dU_{\rm gas} +TdS_{\rm gas} = -dF_{\rm gas}$.

An example of such a system, which is considered in the next section, is an ideal gas in contact with a 1-D chain of oscillators placed in between two thermal reservoirs. In this configuration the chain acts as a reservoir and the mechanical work is done by the pressure exerted by the gas.
	
Thus, we have shown that a class of systems consisting of an ideal gas plus an external system (degree of freedom) that only contributes to $U$ as $TS_{\rm ex}$, allow for a steady state (non divergent) energy flow to be set up between the two thermal reservoirs.\footnote{****}{I do not understand this.}

\ve
 \ce{\it Example system: Oscillator chain connecting thermal reservoirs}

In the previous section we have seen both the necessity to include $S{\rm ex}$, as well as its interpretation. In this section we analyze a mechanical system with one degree of freedom which can add the term  $S_{\rm ex} $ to the ideal gas entropy. A one dimensional chain of harmonic oscillators connecting point particles of equal mass in the presence of an external periodic potential (lattice) is considered. When this chain is placed in contact with an ideal gas, it imposes a constant temperature gradient on the gas.

As shown by Hu, Li and Zhao (20??) using numerical computations, the presence of the external lattice is necessary for Fourier's heat law to be obeyed. In the absence of the lattice the steady state thermal conductivity depends on the total number of masses (oscillators) N while density is held constant. \footnote*{Confusing; what $N$ is this?}This is clearly contrary to the macroscopic phenomena since for example we do not expect the conductivity of a rod to depend on its length. The computations here follow those of Hu et. al and are performed with an external lattice potential.

Consider a chain of point masses connected by oscillators and given by the Hamiltonian:\footnote{**}{Best not to use the letters $U, u$}
$$
H= \sum \Big({p_i^2 \over 2m}+ {k\over2}(X_i-X_{i-1}-a)^2-U(X_i)\Big)
$$
Here, $U(X_i)$ is an arbitrary periodic potential with a spatial period $b$ and $\omega = \sqrt{k/m}$. We can write the Hamiltonian in dimensionless form with scaled parameters:
$$
H= \sum\Big({p_i2\over 2m}+ {k\over m}(x_i-x_{i-1}-c)^2-u(x_i)\Big)
$$
with $c = a/b$ ($a$ is the equilibrium position).

Also, $T_i= <p_i^2>$, where $< >$ denotes the time average. As in Hu et. al the energy current is defined as:
$$
J=  {\p x_i\over\p t} {\p V\over\p x_i}(x_i,x_{i+1}),
$$
where $V$ is the harmonic oscillator potential.

The goal of the numerical computations is to compute the spatial variation of the entropy $S$. Note that $T$ is the only thermodynamic variable defined above in terms of the dynamical quantities. The density does not enter these considerations (this point is discussed further later\footnote {***}{Is it?}). So to find variation of $S$ we use the equation of state relating $S$ and $T$ (in the canonical formalism). For a single harmonic oscillator in contact with a thermal reservoir at temperature $T$, we can derive starting with canonical partition function:
$$
{S\over k}=1+\ln(kT)-\ln{k\over \omega}
$$
Here $k$  is the Boltzmann constant. This equation is obtained by using the quantum formula for the energy oscillator energy levels and then keeping terms to first order in $ {h/\omega}{kT}$ since our treatment is classical. The important point about the above equation is the logarithmic dependence of $S/k$ on $kT$.

 \ve

\ce {\it Calculations}

Fixed boundary conditions are used for the chain. Thermostats at $T_{+}$ and $T_{-}$ are connected to the penultimate masses $(1, n-1)$ on either side. The action of the thermostats is given by the equations:
$
\ddot{x_1} = f_1 - f_2 - \gamma_1\dot{x_1}; \ddot{x_n} = f_{n-1} - f_n - \gamma_n\dot{x_n}
$,
with $f_i = -{\p V(x_i,x_{i-1})}{\p x_i}, V$ is harmonic potential. $\dot{\gamma_1}={(x_1)^2}{T_+}-1$; $\dot{\gamma_n}={(x_n)^2}{T_-}-1$.

This system is solved using 8th order Runge-Kutta algorithm in MATLAB, to find $x_i (t)$. Then the temperature is found using  $T_i= <(p_i)^2>$, where the average is taken over 10000 time units. The results for the spatial profile of kT are plotted below:
\vskip3.5cm

\def\picture #1 by #2 (#3){
  \vbox to #2{
    \hrule width #1 height 0pt depth 0pt
    \vfill
    \special{picture #3}      }
  }
\def\scaledpicture #1 by #2 (#3 scaled #4){{
  \dimen0=#1 \dimen1=#2
  \divide\dimen0 by 1000 \multiply\dimen0 by #4
  \divide\dimen1 by 1000 \multiply\dimen1 by #4
  \picture \dimen0 by \dimen1 (#3 scaled #4)}
  }

\parindent=1pc

\vskip-3cm
 
\epsfxsize.6\hsize
\centerline{\epsfbox{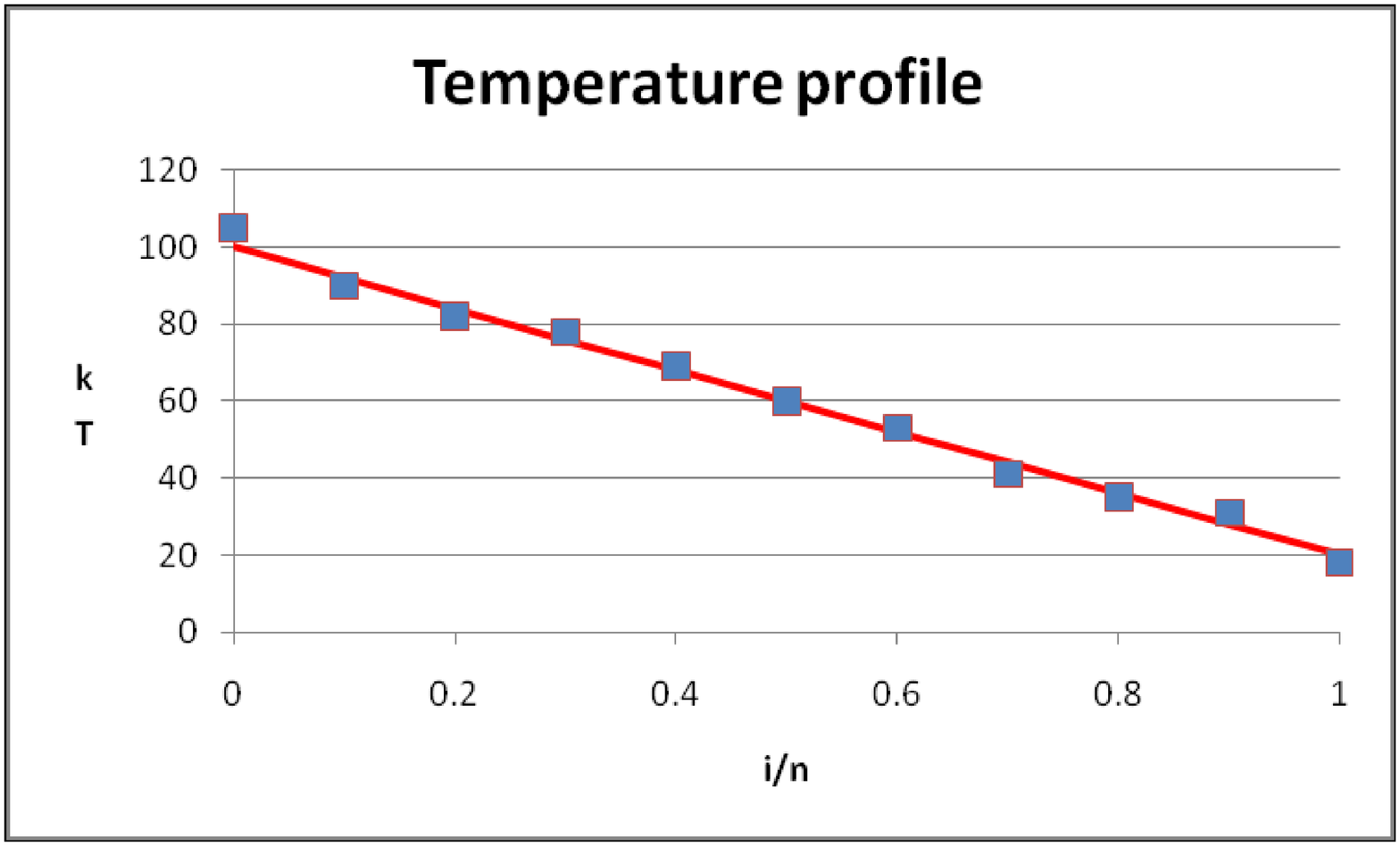}}
\vskip-1cm

\vskip1cm

Figure 1: The vertical axis is $kT$ in scaled units on the x axis, $i$ is the mass (oscillator) number and n is the total number of masses. The red line (slope = -80) was fitted to the computation results
 \b
It is now straightforward to calculate the spatial profile of ${S}/{k}$ using the above result for temperature:
(Only the $\ln(kT)$  term from the equation above for ${S}{k}$ is plotted. The constants can be absorbed into the definition of ${S}/{k})$.
\vskip3.5cm

\def\picture #1 by #2 (#3){
  \vbox to #2{
    \hrule width #1 height 0pt depth 0pt
    \vfill
    \special{picture #3}      }
  }
\def\scaledpicture #1 by #2 (#3 scaled #4){{
  \dimen0=#1 \dimen1=#2
  \divide\dimen0 by 1000 \multiply\dimen0 by #4
  \divide\dimen1 by 1000 \multiply\dimen1 by #4
  \picture \dimen0 by \dimen1 (#3 scaled #4)}
  }

\parindent=1pc

\vskip-3cm
 
\epsfxsize.6\hsize
\centerline{\epsfbox{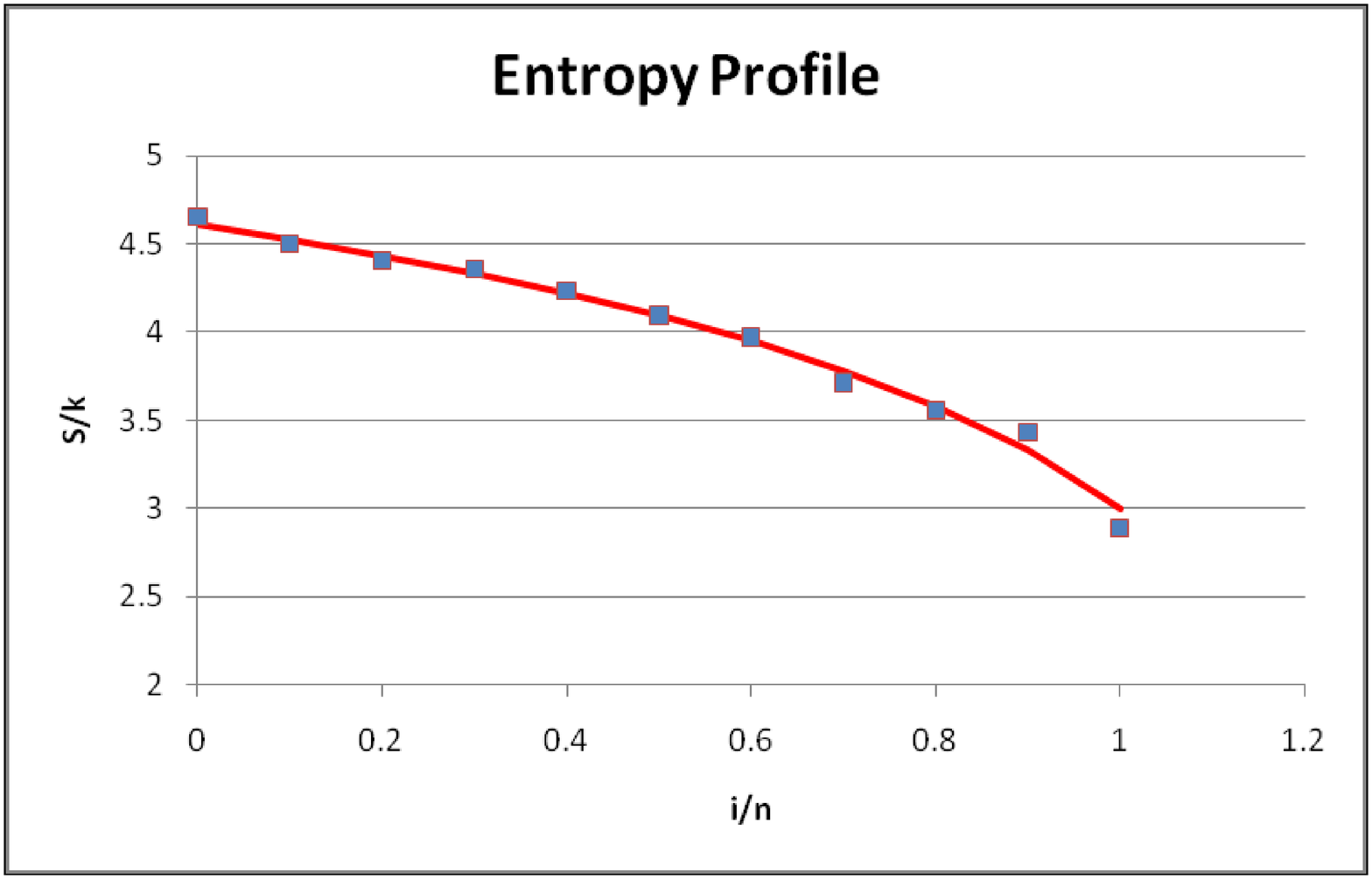}}
\vskip-1cm

\vskip1cm

Figure 2: Entropy Profile. Each point represents the total entropy within bin located at corresponding $i$.

\ve

 \b\b

\no{\bf IV. Second example, the vertical column}

After the initial success of hydro-thermodynamics in accounting for (or at least elegantly summarizing)
the early laboratory experiments, in which the role of the ambient gravitational field could safely be ignored, it is natural to apply the theory to the conditions in the earthly atmosphere. Because this is a complicated system,
in which both radiation and convection play important roles, we shall 
 examine the somewhat academic problem of an isolated, vertical
column at  equilibrium. \footnote*{It would seem natural to study this simplest case first, as a preliminary to developing a theory of atmospheres that are subject to 
additional complications.}

The role of gravity is believed to be understood and defined, if not uniquely, by the equivalence principle of General Relativity: In a non relativistic context it is enough to include a term that represents the gravitational interaction in the hamiltonian,  \footnote
{**}{See for example Eckert (1987)     page  260.  Not the best reference.}  
$$
H =\int d^3x\Big(\phi\rho +\rho\vec v\,^2 +  f(T,\rho) +sT\Big), 
$$
where $\phi = gz$, $z$ the vertical coordinate and $g$ constant.

The variational equations associated with this model are precisely those of the
earliest model atmospheres, with $s = \rho S$ and the specific entropy $S$ taken to be constant.

Variation with respect to $T$ and $\rho$ gives Eq.s (2.2) as before, modified only by the gravitational term, when $\vec v = 0$, 
$$
\R\rho(\ln k - n) + s = 0,~~ {\rm and}~~\phi +\R T(\ln k + 1) + {\p s\over \p \rho}T = c = {\rm constant}.\eqno(4.1)
$$
Multiplying the first equation by $T$ and the other by $\rho$ and subtracting the first from the second we obtain
$$
\rho\phi+[(n+1)\R\rho  + (\rho{\p s\over \p \rho}-s) ] T= c\,\rho.\eqno(4.2)
$$
The situation encountered with the conducting slab is reversed. If $s$ is linear in $\rho$; that is,  if $s = \rho S$, whether or not $S$ is uniform,  then this implies a constant, non zero  lapse rate, ${\rm grad}\,T = $  constant  $\neq 0$. This constant temperature lapse  is consistent with observation in real atmospheres, but it represents a failure in the attempt to describe an isolated column of gas, for it is contradicts  the axiom of uniformity of the temperature - if this axiom is extended so as to apply in the presence of a gravitational field.

Indeed we have been made to understand, or at least to accept, that the equilibrium configuration of an isolated homogeneous substance is isothermal, and with especial emphasis, that this remains true notwithstanding the influence of an ambient gravitational field.

We are thus in an uncomfortable position that we may try to remedy by the same
strategy that was used to account for the boundary conditions associated with slab geometry. That is, we must give up the (admittedly unmotivated) idea that the column is isentropic and instead postulate that it is isothermal, at equilibrium.

We are led to try the  strategy that led to a solution of the slab problem, setting
$$
s = \rho S + S_{\rm ex}.
$$
The equations of motion (4.1 --2) are now
$$
\R \rho(\ln k - n) + \rho S +S_{\rm ex} = 0\eqno(4.3)
$$
and
$$
\phi +\R T(\ln k +1) + ST= c={\rm constant},\eqno(4.4)
$$
combining to give
$$
gz +(n+1)\R T = S_{\rm ex}T/\rho + c,\eqno(4.5)
$$

In this case the pressure is not expected to be uniform. The thermodynamic pressure 
is again equal to $\R \rho T$, but  the hydrostatic condition (1.5) is modified by the
action of gravity
$$
{\rm grad }\, p + \rho\,{\rm grad} \phi=  0.\eqno(1.5')
$$

As in the case of the slab, the thermodynamical relations (4.3) and (4.4)  imply a modification, exactly as in  Eq.(1.16), 
$$
{\rm grad }\, p + \rho\,{\rm grad} \phi = [(1-\rho{\p\over \p \rho})s]\,{\rm grad}\, T-\rho T\,{\rm grad} {
\p s\over \p\rho}.\eqno(4.6)
$$
That $T$ must be uniform is an almost universally accepted axiom. The first term on the right hand side is therefore zero. The hydrostatic condition of the standard theory 
then requires that the second term vanish as well, hence $S= \p s/\p \rho$ is uniform. Then Eq. (4.4) gives the standard result:  the density,  at equilibrium, takes the form
$$
\rho =  \rho_0 =  \e^{-1-S/\R}{T_0}^n \, \e^{(c-
gz)/\R T_0}.\eqno(4.7)
$$
The other equation of motion, Eq.(4.5), tell us that $S_{\rm ex}$ can not be taken to vanish. At equilibrium, on the assumption that ${\rm grad}\, T = 0$,
$$
S_{\rm ex} = {g(z-z_0)\over T_0}\rho_0. \eqno(4.8)
$$
 We can fix the specific entropy $S$ (a constant) and the temperature  (another constant) and calculate the external entropy (variable) and the density (also variable). 
 But that is not how we understand Gibbs'  variational principle, or the two principal relations of thermodynamics. Instead we have been taught to fix the entropy to define the adiabatic system and calculate the extremal of the energy  function with respect to variations of $T$ and $\rho$.  Equations (4.3) and (4.4) are equations of motion to be satisfied by $\rho$ and $T$, for an entropy fixed in advanced; Eq.(4.8) is a constraint
 on the allowed choice of $S_{\rm ex}$ arising from the additional requirement that the temperature at equilibrium be uniform.

This makes it difficult to satisfy the axiom of uniform temperature.  It may be possible to accept the idea that, given time, the entropy will adjust until a uniform temperature is achieved. But this change of entropy is required to take place, for the system under consideration,  only in the presence of gravity.  We thus end up with a modified set of instructions for  incorporating the effect of gravity into the theory:
\b
(1) Add the term $\rho\phi$ to the hamiltonian   and

(2) Add the term $S_{\rm ex}$, precisely as in (4.8),  to the entropy.  
 \b
  \no The influence of gravity is a little more complicated than expected.
  
  It is fortunate that in actual atmospheres, there has been no need to introduce this 
  external contribution to the entropy. The prevailing view is that simplicity is restored
  ($S_{\rm ex}$ is taken to vanish and the atmosphere becomes approximately
   isentropic) through the agency of radiation and convection. The ideal situation, in which gravity acts alone on the isolated gas is seldom considered  theoretically,
   nor is it seen as being worthy of experimental investigation.

\b

Consider now a particular equilibrium state. Choose a value for  $S$ and $T$ and determine the values of the distributions $\rho$ and $S_{\rm ex}$. Having  thus chosen the entropy   we can approach dynamical problems, and especially the propagation of sound on the adiabatic hypothesis. Let us make the common assumption that conduction makes no appreciable contribution, then there is no need to invoke the heat equation. 

Since the propagation of sound implies motion we now include the kinetic energy density $\rho\vec v\,^2$ in the hamiltonian density.
Of the two basic equations (4.3) and (4.4) only the second (the hydrostatic equation)
is affected;  including the kinetic energy and taking the gradient we obtain the Bernoulli equation
$$
{D\vec v\over DT}+{\rm grad }\, p + \rho\,{\rm grad} \phi = S_{\rm ex}\,{\rm grad}\, T-\rho T\,{\rm grad} {
\p s\over \p\rho}.\eqno(5.1)
$$
We have assumed that the right hand side vanishes at equilibrium, and we used the additional requirement of uniform equilibrium temperature to place restrictions on the entropy. But once the entropy has been chosen in accord with this prescription all the equations of motion must continue to hold, and the second term on the right hand side of (5.1) remains zero, but the first term remains.
 
To evaluate the speed of sound we can use any complete set of equations of motion and the simplest set consists of the continuity equation and the two equations  (4.3) and (4.4), the latter modified by the kinetic energy,
$$
\dot\rho + {\rm div}(\rho\vec v) = 0,
$$
$$
\R \rho(\ln k - n) + \rho S +S_{\rm ex} = 0\eqno(4.3)
$$
$$
\dot{\vec v} + {\rm grad}\Big( \phi + \vec v\,^2/2 + \R T(\ln k +1) + ST\Big)={\rm constant},\eqno(4.4')
$$

 The propagation of sound is seen, since first proposed by Laplace, as an adiabatic process, and as a first order perturbation of the state of equilibrium. Without the presence of gravity it leads to the result that the speed of propagation is 
 $
 c = \sqrt{\gamma \R T}.
 $
 In the case of the isolated column the equations that govern the perturbation are, to first order in $\delta \rho$ and $\delta T$ are
 $$
 \delta \dot \rho + {\rm div }(\rho\vec v) = 0,
 $$
 $$
 {\delta k \over k}  = S_{\rm ex} {\delta\rho\over \rho^2},~~{\delta T\over T} = {1\over n}(1-{S_{\rm ex}\over \R \rho}){\delta\rho\over \rho},
    $$
    $$
 \,\dot{\vec v} +  {\rm grad}
\,\delta\Big(\R T(\ln k + 1) + ST\Big) = 0.
 $$
 With the neglect of terms of order $g^2$ we obtain
    $$
  \ddot \rho =  -{\rm div}(\rho \dot {\vec v}) =   \R T {\rm div}\Big(\rho\,{\rm grad}\,[(\gamma- {2g(z-z_o)\over nT}) {\delta \rho\over \rho}]\Big).
   $$
Besides the appearance of damping, we note the dependence of the speed of propagation on the altitude. 

The propagation of sound in  a polytropic atmosphere (taken as a good representative of real atmospheres) is discussed by many authors, se e.g. Stanyukovich (... ).
It appears that there has been no study of the propagation of sound in the ideal
atmosphere (uniform temperature in the presence of the gravitational field).
This may be because the mere statement that $T$ is uniform in the equilibrium configuration is not a dynamical statement that applies to adiabatic changes; it is 
more of the nature of  a boundary condition. In this paper we have found the underlying dynamical statement,  and that puts us in the position of being able to analyze
questions of dynamics.
 
\b\b

\no{\bf VI. Discussion}

The extra contribution $S_{\rm ex}$  was invented, first, to understand the slab.
We expect that it is uncontroversial, although the physical justification may be put into question.  Using only the basic relations of thermodynamics we showed that 
the assumption $s = \rho S$ is inconsistent with the boundary conditions.
The analysis raise an interesting question - panel (2.6):  is the pressure uniform across the slab? This should be easy an easy question to subject to observation.
 
The case of the isolated atmosphere in a gravitational field is different, since a
physical interpretation of the extra piece in the entropy is lacking. Indeed, the
physical model that was invoked in the case of the slab is much less convincing 
in the case of the atmosphere. We have attempted to incorporate the assumption of  a uniform temperature into adiabatic thermodynamics. If the attempt is found to be
unconvincing then the issue remains open. We believe that it deserves to be debated and that, perhaps, after all,  the question is worthy of experimental investigation.

\ve

\no{\steptwo Acknowledgements}

C.F. Thanks J. Rudnick and G. Williams for helpful discussions.  A.P. thanks J.Rudnick
for financial support.

\bb
{\steptwo References}

\no Callen, H.B., {\it Thermodynamics and an Introduction to Thermostatistics}, 

John Wiley \&  Sons, N.Y. 1985. enter page number on page 3

\no Eckert, E.R.G., {\it Heat and Mass Transfer} 1959.  

\no Hu, B.,   Li, B. and  Zhao,  H., "Heat Conduction in one-dimensional chains",  

Phys. Rev E. Vol. 57, 3 (1999).

\no Laplace,  P.S., {\it Trait\'ee de Mechanique Celeste}, Paris 1825.

\no Poisson, , S.D., {\it Th\'eorie math\'ematique de la chaleur},  1835.

\no Stanyukovich, K.P., {\it Unsteady motion of continuous media},
Pergamon Press N. Y. 1960.

 \end